\documentclass[12pt,a4paper,twoside]{article}
\usepackage{graphicx,cite}

\usepackage{epsfig}
\usepackage{graphicx}
\usepackage{multicol}
\usepackage{rotating}

\usepackage{color}

\def\be{\begin{equation}}
\def\ee{\end{equation}}
\def\bea{\begin{eqnarray}}
\def\eea{\end{eqnarray}}

\def\bbuildrel#1_#2^#3{\mathrel{\mathop{\kern 0pt#1}\limits_{#2}^{#3}}}
\def\slash#1{\setbox0=\hbox{$#1$}#1\hskip-\wd0\dimen0=5pt\advance
       \dimen0 by-\ht0\advance\dimen0 by\dp0\lower0.5\dimen0\hbox
         to\wd0{\hss\sl/\/\hss}}

\newcommand{\gae}{\lower 2pt \hbox{$\, \buildrel {\scriptstyle >}\over
    {\scriptstyle \sim}\,$}}
\newcommand{\lae}{\lower 2pt \hbox{$\, \buildrel {\scriptstyle <}\over
    {\scriptstyle \sim}\,$}}

\newcommand {\fig} [1] {fig.~\ref{fig:#1}}

\def\gsim{\raise0.3ex\hbox{$\;>$\kern-0.75em\raise-1.1ex\hbox{$\sim\;$}}}
\def\lsim{\raise0.3ex\hbox{$\;<$\kern-0.75em\raise-1.1ex\hbox{$\sim\;$}}}

\begin{document}

\begin{titlepage}

\begin{flushright}
CERN-PH-TH/2008-203\\
SLAC-PUB-13424\\
hep-ph/yymmnnn\\[2cm]
\end{flushright}

\begin{center}
\setlength {\baselineskip}{0.3in}

{\bf\large Flavour violating squark and gluino decays}

\vspace{0.5cm}

\setlength {\baselineskip}{0.2in} {\large Tobias Hurth$^{1,2}$, 
Werner Porod$^{3}$}\\[8mm]

$^1$~{\it CERN, Dept. of Physics, Theory Division, CH-1211 Geneva, Switzerland.}\\[5mm]

$^2$~{\it SLAC, Stanford University, Stanford, CA 94309, USA}\\[5mm]
 
$^3$~{\it Institut f\"ur Theoretische Physik und Astrophysik 
Univ. W\"urzburg, D-97074 W\"urzburg, Germany}\\[5mm]

{\bf Abstract}\\[5mm]
\end{center} 
\setlength{\baselineskip}{0.2in} 
We consider scenarios with large flavour violating
entries in the squark mass matrices focusing on the mixing between
second and third generation squarks. These entries govern both,
flavour violating low energy observables on the one hand and
squark and gluino decays on the other hand.
We first discuss the constraints on the parameter space due to the
recent data on $B$ mesons from the $B$ factories and Tevatron. We then
consider flavour violating squark and gluino decays and show that they
can still be typically  of order 10\% despite the stringent
constraints from low energy data.  Finally we briefly comment on the
impact for searches and parameter determinations at future collider
experiments such as the upcoming LHC or a future International Linear
Collider.

\end{titlepage}

\section{Introduction} 
Theoretical arguments like the hierarchy problem motivate the general
expectation that the experiments at the LHC will 
lead to discoveries of new  degrees  of freedom at the TeV energy scale. 
The precise nature of this new physics is unknown, but it most probably
will  answer some of the fundamental questions 
related to the origin of electroweak symmetry breaking.

Rare $B$ and kaon decays (for a review see
\cite{Hurth:2007xa,Hurth:2003vb}) representing loop-induced processes
are highly sensitive probes for new degrees of freedom beyond the SM
establishing an alternative way to search for new physics. 
However, this indirect search for new physics signatures
within flavour physics takes place today in complete darkness, given
that we presently have no direct evidence of new particles beyond the
Standard Model (SM).  But the day the existence of new degrees of
freedom is established by the Large Hadron Collider (LHC), the
searches for anomalous phenomena in the flavour sector will become
mandatory.  The problem then will no longer be to discover new
physics, but to measure its (flavour) properties.

Thus, within the next decade an important interplay of flavour and
high-$p_T$  physics most probably will take place. For example, within
supersymmetric extensions of the SM, the measurement of the flavour
structure is directly linked to the crucial question of the
supersymmetry-breaking mechanism as the soft SUSY breaking terms are
the source of flavour structures beyond the SM.  LHC has the potential 
to  discover
strongly interacting supersymmetric particles up to a scale of 2 TeV
and to  measure several of their properties 
\cite{Ball:2007zza,delAguila:2008iz,Buchalla:2008jp,Aad:2009wy}. 
This information can be used for a refined analysis of flavour physics
observables indicating possible flavour structures and, thus, give
important information for distinguishing between models of
supersymmetry breaking.

Data from $K$ and $B_d$ physics show  that new sources of flavour violation
in \mbox{$s \rightarrow d$} and \mbox{$b \rightarrow d$} 
are strongly constrained, while 
the possibility of sizable  new 
contributions to \mbox{$b \rightarrow s$}  
remains open. We also have hints from 
model building: flavour models are not very effective in  constraining 
the \mbox{$b\rightarrow s$}  sector~\cite{Masiero:2001cc}. 
Moreover, in SUSY-GUTs 
the large mixing angle in the neutrino sector relates to  large mixing in 
the right-handed $b$-$s$ sector~\cite{Moroi:2000tk,Chang:2002mq,Harnik:2002vs}.

As we explicitly show in this paper, such flavour information on the
$b \rightarrow s$ observables is complementary to the high-$p_T$ data
of the LHC.  In fact,  squark and gluino decays are governed by the same
mixing matrices as the contributions to flavour violating loop
transitions.  This allows for possible direct correlations between
flavour non-diagonal observables in $B$ and high-$p_T$ physics. We
already anticipate that the present bounds on squark mixing, induced
by the low-energy data on $b \to s$ transitions, still allow for large
contributions to flavour violating squark decays at tree level.  Due
to the restrictions in flavour tagging at the LHC, additional
information from future flavour experiments will be necessary to
interpret those LHC data properly.  Also the measurement of
correlations between various squark decay modes at a future ILC would
provide information about the flavour violating parameters.

Flavour violating squark and gluino decays have already been 
considered some time ago \cite{Hurth:2003th}. In the present paper we
generalize and update the previous analysis including additional
experimental constraints.  More recently, similar work on the charged
Higgs boson production was presented; it was shown that squark mixing
can significantly change the production pattern
\cite{Dittmaier:2007uw}. In another work the interesting question was
addressed if high-$p_T$ data can contribute to the solution of the
flavour problem \cite{Grossman:2007bd,Feng:2007ke}.  Recently, it was
shown that one can also derive significant bounds on flavour-violating
parameters of the squark sector by requiring that the radiative
corrections to the CKM elements do not exceed the experimental
values\cite{Crivellin:2008mq}. Flavour violating squark production has
been studied in \cite{Bozzi:2005sy,Bozzi:2007me,Fuks:2008ab} where it
has been found that flavour violating production can be sizable after
taking into account constraints from the $b \rightarrow s$ observables
and the anomalous magnetic moment of the muon.

The paper is organized as follows:
In Chapter 2 we review the necessary information about 
flavour mixing in the MSSM and about decays of squarks and gluinos
in order to introduce our notation. In Chapter 3 
we present our phenomenological  results and in Chapter 4 we discuss 
their impact for LHC and ILC.

\section{Preliminaries}
\subsection{Decays of squarks and gluinos}
In the study of squark decays two scenarios can be distinguished
depending on the SUSY spectrum:
\begin{enumerate}
\item $m_{\tilde g} > m_{\tilde q_i}$ ($q=d,u$; $i=1,\dots,6$):  In 
  this case the gluino will mainly decay according to
  \begin{eqnarray}
    \tilde g \to d_j \, \tilde d_i \,,\,\,\,    \tilde g \to u_j \, \tilde u_i 
  \end{eqnarray}
  with $d_j = (d,s,b)$ and $u_j = (u,c,t)$ followed by squark decays
  into neutralino and charginos
  \begin{eqnarray}
   \tilde u_i \to u_j \tilde \chi^0_k \, , \,  d_j \tilde \chi^+_l \,, \,\,\,  
   \tilde d_i \to d_j \tilde \chi^0_k \, , \,  u_j \tilde \chi^-_l \,\,.
   \end{eqnarray}
  In addition there can be decays into gauge  and Higgs bosons if
  kinematically allowed: 
  \begin{equation}
   \tilde u_i \to Z \tilde u_k\,, \,\,  H^0_r \tilde u_k\,, \,\,
       W^+ \tilde d_j\,, \,\,  H^+ \tilde d_j;\,\,\,\,\,\,\,\,\,   
   \tilde d_i \to  Z \tilde d_k\,, \,\,  H^0_r \tilde d_k\,, \,\,
       W^- \tilde u_j\,, \,\,  H^- \tilde u_j
  \end{equation}
  where $H^0_r = (h^0, H^0, A^0)$, $k < i$, $j=1,\dots,6$.
  Note, that due to the fact  that there is left-right mixing in the
  sfermion mixing, one has flavour changing neutral decays into $Z$-bosons
  at tree-level. 
\item $m_{\tilde g} < m_{\tilde q_i}$ ($q=d,u$; $i=1,\dots,6$): In 
  this case the squarks decay mainly into a gluino,
  \begin{eqnarray}
   \tilde u_i \to u_j \tilde g \,, \,\,\,\,\,\,
   \tilde d_i \to d_j \tilde g \,\,
  \end{eqnarray}
  and the gluino decays via three-body decays and loop-induced two-body
  decays into charginos and neutralinos
  \begin{eqnarray}
    \tilde g \to d_j \, d_i \, \tilde \chi^0_k \,,\,\,
                   u_j \, u_i \, \tilde \chi^0_k \,,\,\,\,\,\,\,
    \tilde g \to u_j \, d_i \, \tilde \chi^\pm_l \,,\,\,\,\,\,\,
    \tilde g \to g  \, \tilde \chi^0_k
  \end{eqnarray}
  with $i,j=1,2,3$, $l=1,2$ and $k=1,2,3,4$. Note that the first two decay
  modes may
contain 
  states with quarks of different generations of quarks.
\end{enumerate}
Explicit formulas for the partial widths including flavour effects can
be found in Ref. \cite{Bozzi:2007me}.  The flavour mixing final states of
the decays listed above are constrained by the fact that all observed
phenomena in rare meson decays are consistent with the SM predictions.
As we will show in the next sections, there are regions in the
parameter space where the flavour violating decay modes can be even of
the order of 10\%.

\subsection{Flavour changing neutral currents}
Within the Minimal Supersymmetric Standard Model (MSSM) there are two
new sources of flavour changing neutral currents (FCNC), namely new
contributions which are induced through the quark mixing as  in the
SM and generic supersymmetric contributions through the squark
mixing.  The latter  is described by their mass matrices,  
\begin{eqnarray}
{\cal M}_f^2 \equiv  \left( \begin{array}{cc}
  M^2_{\,f,\,LL} +F_{f\,LL} +D_{f\,LL}           & 
                 M_{\,f,\,LR}^2 + F_{f\,LR} 
                                                     \\[1.01ex]
 \left(M_{\,f,\,LR}^{2}\right)^{\dagger} + F_{f\,RL}^* &
             \ \ M^2_{\,f,\,RR} + F_{f\,RR} +D_{f\,RR}                
 \end{array} \right)
\label{massmatrixD}
\end{eqnarray}
where $f$ stands for up- or down-type squarks. 
In the super-CKM basis in which the quark mass matrix is diagonal they
read as
\begin{eqnarray}
D_{f\,LL} = (T_{3,f} - e_f \sin^2\theta_W) \cos(2 \beta)  m^2_Z  \,\,,\,\,\,\,\,\,
D_{f\,RR} =  e_f \sin^2\theta_W \cos(2 \beta)  m^2_Z
\end{eqnarray}
for the $D$-terms,
\begin{equation}
F_{f\,LL,ij} = F_{f\,RR,ij} = m^2_i \delta_{ij}\,\,,\,\,\,\,\,\,  
F_{f\,RL,ij} =  - \mu m_i \delta_{ij} (\tan\beta)^{-2 T_{3,f}}
\end{equation}
where $m_i$ are the corresponding quark masses, $e_f$ the electric charge,
and $T_{3,f}$ the weak isospin of the corresponding left-handed quark. In
this basis the $F$- and $D$-terms are flavour diagonal and all flavour
violation beyond the CKM resides in the soft SUSY breaking terms:
\begin{eqnarray} \label{Mmatrices1}
M^2_{d,LL} &=& V_{CKM} ^\dagger     M^2_{u,LL} V_{CKM}   =    {\hat m_{\tilde Q}}^2 \equiv V^\dagger_d \,m^2_{\tilde Q}\, V_d\,,~~~        \\
M^2_{d,RR} &=&   {\hat m_{\tilde d}}^2 \equiv
 U^\dagger_d \,{m^2_{\tilde d}}^T\, U_d\,          \,\, , 
\hspace{3cm} M^2_{u,RR} =  {\hat m_{\tilde u}}^2 \equiv U^\dagger_u \,
{m^2_{\tilde u}}^T\, U_u\,,~~~       \\
M^2_{d,LR} &=& v_1/ \sqrt{2}\,      {\hat T_{D}} \equiv     v_1/ \sqrt{2}\,         
U^\dagger_d \,T_{D}^T\, V_d 
\,\, , \hspace{1cm} M^2_{u,LR} = v_2 /\sqrt{2} \,   {\hat T_{U}} \equiv             
v_2 /\sqrt{2} \,   U^\dagger_u \,T_{U}^T\, V_u \,,  \label{Mmatrices}
\end{eqnarray}
where the un-hatted mass matrices $m^2_{Q,u,d}$ and trilinear
interaction matrices $T_{U,D}$ are given in the electroweak basis.
The transformations $V_{u,d}$ and $U_{u,d}$ just bring the quarks from
the interaction eigenstate basis to their mass eigenstate basis, so
$V_{{CKM}}=V_u^\dagger V_d$.  The relation between $M_{u,LL}$ and
$M_{d,LL}$ is due to $SU(2)_L$ gauge invariance.  The $T$-matrices are
in general non-hermitian.\footnote{We follow here the conventions of
  the SUSY Les Houches Accord, for more details see Ref.\cite{LesHouches}.}

These additional flavour structures induce flavour violating couplings
to the charginos, neutralinos and gluinos in the mass eigenbasis,
which give rise to additional contributions to observables in the $K$
and $B$ meson sector.
The low-energy observables can thus be used to
constrain the size of the off-diagonal elements of the mass matrices $
M^2_{\,f,\,LL}$, $ M^2_{\,f,\,RR}$, and $M^2_{\,f,\,LR}$. Some of the
flavour-violating terms may turn out to be poorly constrained. Thus,
it is suitable to rely on the mass eigenstate formalism, which remains
valid -- in contrast to the mass insertion approximation -- even when
the intergenerational mixing elements are large. The diagonalization
of the two $6 \times 6$ squark mass matrices squared ${\cal M}^2_d$
and ${\cal M}^2_u$ yields the eigenvalues $m_{\tilde{d}_k}^2$ and
$m_{\tilde{u}_k}^2$ ($k=1,...,6$). The corresponding mixing 
matrices $R^f_{ij}$ relate the mass eigenstates  with the
electroweak eigenstates
\begin{equation}
\tilde u_k = R^u_{kj} \tilde u^{ew}_j \,\,,\,\, \tilde d_k = R^d_{kj} \tilde d^{ew}_j
\end{equation}
where 
$\tilde u^{ew}_j \in \{\tilde u_L,\tilde c_L,\tilde t_L,\tilde u_R,\tilde c_R,\tilde t_R\}$ and 
$\tilde d^{ew}_j \in \{\tilde d_L,\tilde s_L,\tilde b_L,\tilde d_R,\tilde s_R,\tilde b_R\}$,
 respectively.

As usual, the flavour off-diagonal elements of the squark mass
matrices are normalized by the diagonal ones. One uses  
the average of the diagonal elements (trace of the
mass matrix divided by six) in the up and down sector, denoted by
$m_{\tilde{q}}^2$. The observables can then be studied as a function
of the normalized off-diagonal elements
\begin{equation} 
\delta_{LL,ij} = \frac{(M^2_{\,f,\,LL})_{ij}}{m^2_{\tilde{q}}}\,, 
\hspace{1.0truecm}
\delta_{f,RR,ij} = \frac{(M^2_{\,f,\,RR})_{ij}}{m^2_{\tilde{f}}}\,, 
\hspace{1.0truecm}
(i \ne j) 
\label{deltadefa}
\end{equation}
\begin{equation} 
\delta_{f,LR,ij} = \frac{(M^2_{\,f,\,LR})_{ij}}{m^2_{\tilde{f}}}\,,
\hspace{1.0truecm}
\delta_{f,RL,ij} = \frac{(M^2_{\,f,\,RL})^\dagger_{ij}}{m^2_{\tilde{f}}}\,.
\phantom{\hspace{1.0truecm}
(i \ne j)} 
\label{deltadefb}
\end{equation}
where $f$ is either $u$ or $d$ for $u$-squarks and $d$-squarks, respectively.
We emphasize that  a consistent analysis of the bounds  should 
also include interference effects between the various contributions, 
namely the interplay between the various sources of flavour violation 
and the interference effects  of SM and various new-physics contributions.
(see Ref.\cite{Besmer:2001cj}).

At present, new physics contributions to $s \to d$ and $b\to d$
transitions are strongly constrained.  In particular, the transitions
between first- and second-generation quarks, namely FCNC processes in
the $K$ system, lead to very strong constraints on the parameter space
of various new physics models.  However, most of the phenomena
involving $b \rightarrow s$ transitions are still largely unexplored
and leave open the possibility of large new physics effects, in spite
of the strong bounds of the three most important $b \rightarrow s$
observables, namely, the inclusive decay modes $\bar B \rightarrow X_s
\gamma$ and $\bar B \rightarrow X_s
\ell^+\ell^-$ and the $B_s$ -- $\bar B_s$ mixing.
It is well-known that the gluino contributions to these two inclusive
decay modes are mostly sensitive to $\delta_{d,23,LR}$ and
$\delta_{d,23,RL}$ in the leading order approximation
\cite{Ciuchini:2002uv,Ciuchini:2007ha,Ciuchini:2006dx}. However, the
restriction to a single $\delta$ parameter and to the gluino
contribution lead to unrealistically strong constraints as was already
pointed out in Ref. \cite{Besmer:2001cj}. There are also sensitivity
to combinations of other $\delta$ parameters and, moreover, depending
on the precise choice of the supersymmetric flavour- diagonal
parameters, other supersymmetric contributions which are also
sensitive up-squark mixing parameters. The gluino contributions to
$B_s-\bar{B}_s$ mixing are mainly sensitive to the combination
$\delta_{LL} \delta_{d,RR} $ and $\delta_{d,LR} \delta_{d,RL}$.  The
latter is not really relevant due to the constraints already induced
by the inclusive decay  modes. But also here sensitivity of that
observable to other squark mixing parameters are not negligible.

Moreover, there is an impact of squark mixing including the third
generation on the lightest Higgs mass and the $\rho$ parameter which
also includes the $\delta$ parameter in the up-squark sector (see for
example \cite{Heinemeyer:2004by}).  There are the Tevatron bounds on
squark masses and certain constraints from dark matter
phenomenology. Also the diagonal $T$ entries (see Eq.~\ref{Mmatrices})
have to be tuned that they lead to perturbatively quark
masses. Finally, we recall that there are also constraints on some
$\delta$ parameters from theoretical considerations, namely that one
should avoid colour and charge breaking minima \cite{casas}.  All
resulting numerical bounds and constraints will be listed below.

\section{Phenomenological analysis}
\subsection{Benchmark points, experimental and theoretical constraints}

In our present analysis, we first fix the flavour-diagonal set of
parameters and then we vary the flavour-nondiagonal parameters and
explore the bounds on those parameters by theoretical and experimental
constraints.  We fix the flavour-diagonal parameter set following
three very popular SUSY benchmark points, namely SPS1a'
\cite{Aguilar-Saavedra:2005pw} ,$I'{}'$ and $\gamma$
\cite{DeRoeck:2005bw}. The first two points are mSUGRA points whereas the third
one is a mSUGRA point with non-universal Higgs-mass parameters at the
GUT-scale.  SPS1a' contains the lightest spectrum with squarks around
500 GeV and $m_{\tilde g}$ around 600 GeV, $\tan\beta=10$, followed by
$\gamma$ with squark masses around 600 GeV, $m_{\tilde g}$ around 580
GeV, $\tan\beta=20$, and $I'{}'$\,\, with squark masses around 730
GeV, $m_{\tilde g}$ around 1000 GeV, $\tan\beta=35$.  For more
information on the benchmark points see
Refs.~\cite{Aguilar-Saavedra:2005pw,DeRoeck:2005bw}. All of them are
consistent with WMAP data \cite{Spergel:2006hy} and measurements of
the anomalous magnetic moment of the muon.

On the  flavour-nondiagonal parameter set we pose the following concrete 
constraints:
\begin{itemize}
\item  We explicitly check if  all our data points fulfill the theoretical vacuum
stability constraints~\cite{casas}.
Actually, all our data points fulfill the conservative condition that in the LR and RL 
submatrix off-diagonal  elements are not larger than diagonal elements, which implies
the usual vacuum stability conditions \cite{casas}.
\item Among the constraints from electroweak precision data, the most important one is 
$m_{h^0} \ge 114.4$ GeV, where we  add 3 GeV  to the theoretical prediction of $m_{h^0}$
as a measure of the theoretical uncertainty~ \cite{Degrassi:2002fi,Heinemeyer:2004gx,Allanach:2004rh}. Furthermore we
 require that $M_W = 80.40\pm 0.03$~GeV where we take as input $m_Z=91.187$~GeV,
$G_F=1.16639 \cdot 10^{-5}$ GeV$^{2}$ and $\alpha_{em}(0) = 1/137.0359895$~\cite{Heinemeyer:2006px}.
\item Squark Tevatron bounds on squarks are of the order of 250 GeV  
depending on the SUSY spectrum \cite{Amsler:2008zz}.
\item The explicit experimental constraints from the  most important
 flavour observables we use in our analysis   are 
\cite{Misiak:2006zs,Lunghi:2006uf,Cho:1996we,Hurth:2003dk,Buras:2002vd,Huber:2007vv}:
\begin{eqnarray}
2.67 <& Br(\bar B \to X_s \gamma)\times 10^{4} &< 4.29\\
13.5  <& \Delta M_{B_S}\,\, ps   &< 21.1 \\
1.05  <&  BR(\bar B \to  X_s l^+l^-)_{\mbox{low} q^2} \times 10^{6} &<  2.15 \\
&  BR(B_s \to \mu^+ \mu^-) \times 10^{8} &\le  5.8 
\end{eqnarray}
Those bounds include experimental {\it and} theoretical errors which
are linearly added.  Explicitly our bounds are the experimental $95\%$
bounds where twice the SM error is added in order to take into account
uncertainties of the new physics contributions in a conservative
way. We have also checked that the recent experimental data on $B \to
\tau \nu$ do not give additional constraints.

\item Another requirement is that the lightest stable particle (LSP) should be neutral
(but not necessarily be the neutralino) which however is fulfilled once
the experimental bounds on the squark masses are taken into account.
\end{itemize} 

For the numerical evaluation we use an updated version of {\tt SPheno}
\cite{Porod:2003um} which has been extended to accept flavour mixing
entries in the sfermion mass matrices.  The masses and mixings of all
SUSY particles are calculated at the one-loop level where the formulas
of ref.~\cite{Pierce:1996zz} have been extended for the flavour-mixing
case \cite{werner2}.  The masses of the neutral Higgs bosons are
calculated at the two-loop level.  For the $B$-physics observables we
use the formulas of
Refs.~\cite{Lunghi:2006hc,Misiak:2006zs,Bobeth:2001jm,Baek:2001kh} for
$\bar B \to X_s \gamma$,
Refs.~\cite{Huber:2007vv,Huber:2005ig,Bobeth:2001jm,Baek:2001kh} for
$\bar B \to X_s \ell^+ \ell^-$, and Ref.~\cite{Buras:2002vd} for
$\Delta M_{B_s}$ and $B_s \to\mu^+\mu^-$.  The branching ratios for
squark and gluino decays are calculated using tree-level accuracy but
with running couplings evaluated at the  scale $Q=m$ where $m$ is
the mass of the decaying particle.

\subsection{Results}

\subsubsection{Constraints on flavour parameters}

In our analysis we want to identify the flavour violating 
decay channels of squarks and the gluino with potentially large 
branching ratios.
And we  are interested  
how the various decays can be combined to obtain information on the
parameters. 

In Table \ref{tab:bounds} we present the allowed ranges of the
flavour-diagonal parameters $\delta_{ij}$ in the $b$-$s$ sector using
the so-called `one-mass-insertion approximation' where just one
$\delta$ parameter is assumed to be active and all the others are set
to zero. Note that we have calculated observables using the exact
diagonalization of the mass matrix of eq.~(\ref{massmatrixD}).
\begin{table}[t]
\caption{Allowed ranges of the $\delta$ parameters  in the neighborhood of the benchmark points  SPS1a', 
$\gamma$ and I${}'{}'$ taking into account the experimental
information given in the text and assuming that only one
flavour-mixing parameter is present. The regions correspond to 95\%
CL.}
\label{tab:bounds}
\begin{center}
\begin{tabular}{|c|c|c|c|}
\hline
 & SPS1a' & $\gamma$ & I${}'{}'$\\
\hline
$\delta_{LL,23}$ & (-0.05,0.03) & (-0.037,0.005) & (-0.06,0.001) \\ \hline
$\delta_{d,RR,23}$ & (-0.43,0.66) & (-0.29,0.48) & (-0.5,0.45)  \\ \hline
$\delta_{u,RR,23}$ & (-0.7,0.7) & (-0.54,0.43) & (-0.55,0.45) \\ \hline
$\delta_{u,LR,23}$ & (-0.16,0.08)& (-0.16,0.06) & (-0.35,0.05)\\ \hline
$\delta_{u,LR,32}$ & (-0.7,0.54) & (-0.5,0.2) & (-0.7,0.27) \\ \hline
$\delta_{d,LR,23}$ & (-0.0047,0.0046) & (-0.006,0.001) & (-0.01,0.0015)\\ \hline
$\delta_{d,LR,32}$ & (-0.019,0.02)  & (-0.015,0.015) &(-0.004,0.003) \\ \hline
\end{tabular}
\end{center}
\end{table}
Our findings are consistent with the results
of previous analyses in Refs.\
\cite{Besmer:2001cj,Ciuchini:2002uv,Cao:2006xb,Cao:2007dk,Dittmaier:2007uw}, 
in particular the pattern of the constraints on the 
up-quark sector. Here it turns out that the parameter $\delta_{u,LR,23}$
receives bounds from the flavour and electroweak constraints while the other 
two parameters $\delta_{u,LR,32}$ and $\delta_{u,RR,23}$ are unconstrained in scenarios
with low and moderate $\tan\beta$ and weakly constrainted once $\tan\beta$ gets large.

\begin{figure}[t]
 \unitlength 1mm
\begin{picture}(170,75)
\put(10,3){\mbox{\epsfig{figure=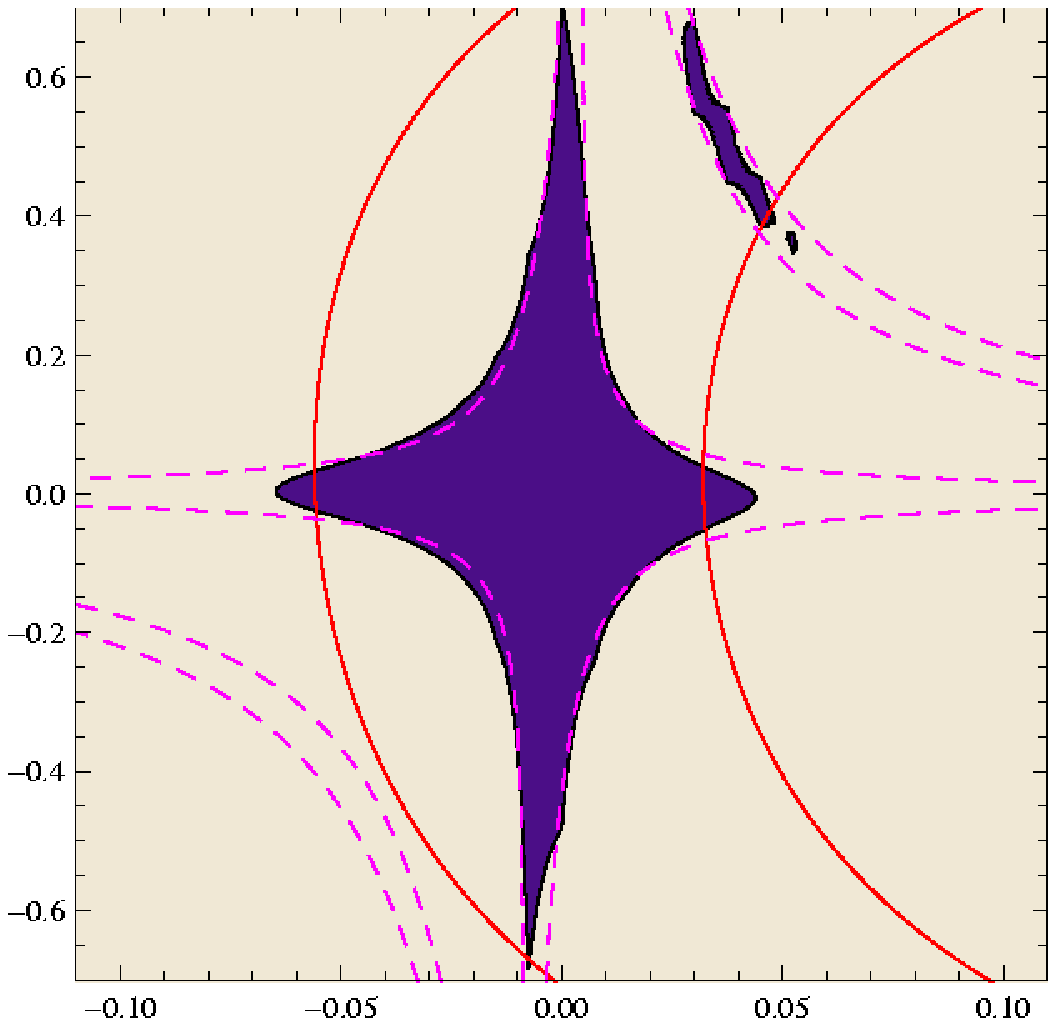,height=7cm,width=7cm}}}
\put(42,-2){\mbox{$\delta_{LL,23}$}}
\put(6,33){\begin{rotate}{90}\mbox{$\delta_{d,RR,23}$}\end{rotate}}
\put(7,75){\mbox{\bf a)}}
\put(96,3){\mbox{\epsfig{figure=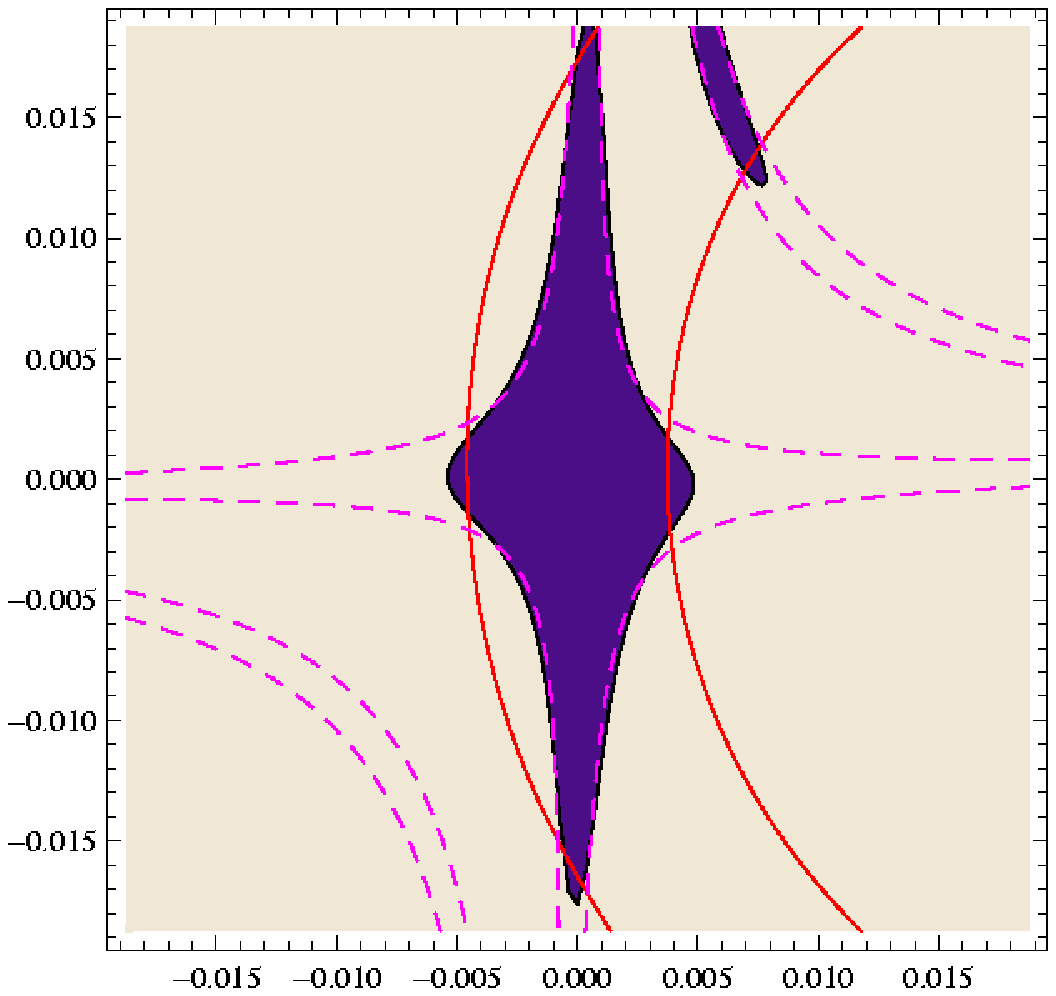,height=7cm,width=7cm}}}
\put(127,-2){\mbox{$\delta_{D,LR,23}$}}
\put(92,35){\begin{rotate}{90}\mbox{$\delta_{D,LR,32}$}\end{rotate}}
\put(93,75){\mbox{\bf b)}}
\end{picture}
\caption{Allowed regions  a) in the $\delta_{d,RR,23}$-$\delta_{LL,23}$  and  b) in the
$\delta_{D,LR,23}$-$\delta_{D,LR,32}$ plane for SPS1a'. All other than the shown
flavour off-diagonal elements are put to zero.
The lines correspond to: full red lines $b\to s \gamma\,,\, 2.67 10^{-4}$ and ,
 $b\to s \gamma\,,\, 4.29 10^{-4}$, dashed magenta  lines
 $|\Delta_{M_{B_s}}| = 13.5$ ps$^{-1}$,
and  $|\Delta_{M_{B_s}}| = 21.1$ ps$^{-1}$. The blue (dark) area shows the regions
consistent with all data at 95\% CL.}
\label{fig:2dregions}
\end{figure}

In Figure \ref{fig:2dregions} we present regions in the 
$\delta_{d,RR,23}$-$\delta_{LL,23}$ (left) and in the
$\delta_{d,LR,23}$-$\delta_{d,LR,32}$ planes (right) consistent with experimental
data for the SPS1a' benchmark point. 
We show here the lines corresponding to the constraints from $b\to s\gamma$ using red (full)  lines and $\Delta M_{B_s}$ using magenta (dashed) ones. The areas  consistent with all
constraints correspond  to the blue (dark) areas. In both cases the complete
plane is compatible also  with the constraints due to the observable 
$b\to s l^+ l^-$, $B_s \to \mu^+ \mu^-$ and $B_u \to \tau \nu$.

One sees the regions allowed by all constraints
corresponding to the blue (dark) area are mainly along the axes as one
would naively expect. Note, however, that the constraint on a
parameter gets more involved once one allows for additional sources of
flavour violation.  For example,  Figure \ref{fig:2dregions}a shows
that for small negative $\delta_{LL,23}$ the constraint on
$\delta_{d,RR,23}$ becomes first weaker; for larger $\delta_{LL,23}$ the bound
is      stronger again.

In both planes of Figure 1 we find an additional
region consistent with the data in the upper right corner
where both flavour mixing parameters
are sizable. However, in these regions clearly some cancellations between
various contributions to the  observable take place. In
particular the gluino and the chargino contributions for $\Delta
M_{B_s}$ are large and of opposite sign.  Their modulus can reach
about twice the size of the SM contribution.  

The other two benchmark
points show very similar features but for the fact that in case of
$I'{}'$ also $B_s \to \mu^+ \mu^-$ would exclude part of the allowed
regions.

\subsubsection{Flavour violating decays of squarks and gluinos}
\label{sec:susydecays}

\begin{table}[ht] 
\caption{Branching ratios larger than 1\% for  two study points. The flavour
diagonal entries are according to SPS1a'. $\tilde u_i$ decays are like in SPS1a'
\cite{Aguilar-Saavedra:2005pw} and in both scenarios 
BR$(\tilde d_3 \to \tilde \chi^0_1 d) = 99.1$\%.}
\label{tab:BR}
%\begin{center}
\vskip 2mm \hskip6mm
\begin{tabular}{|c||r|r|r||r|r|r|}
\hline
decaying & \multicolumn{6}{|c|}{final states and corresponding branching ratios in \% for.}  \\
 particle & \multicolumn{3}{|c||}{I. $\delta_{LL,23}=0.01,\delta_{D,RR23}=0.1$}
        & \multicolumn{3}{|c|}{II. $\delta_{LL,23}=0.04,\delta_{D,RR23}=0.45$}
\\ \hline
$\tilde d_1 \to $  
   & $\tilde \chi^0_1 b$\,,\, 4.4 & $\tilde \chi^0_2 b$\,,\, 29.8 &  $\tilde \chi^-_1 t$\,,\, 37.0
   & $\tilde \chi^0_1 s$\,,\, 36.8 & $\tilde \chi^0_1 b$\,,\, 42.2 &   $\tilde \chi^0_2 b$\,,\, 10.9
   \\ 
   & $\tilde u_1 W^-$\,,\,  27.7 & &
   &  $\tilde \chi^-_1 t$\,,\, 9.6 & & \\
\hline
$\tilde d_2 \to $  
   & $\tilde \chi^0_1 s$\,,\, 8.0 &  $\tilde \chi^0_1 b$\,,\, 6.4 &  $\tilde \chi^0_2 b$\,,\, 19.0
   & $\tilde \chi^0_1 b$\,,\, 2.1 & $\tilde \chi^0_2 b$\,,\, 27.3 &  $\tilde \chi^-_1 t$\,,\, 34.6
   \\ 
   & $\tilde \chi^0_3 b$ \,,\, 1.1 & $\tilde \chi^0_4 b$ \,,\, 1.8 & $\tilde \chi^-_1 t$\,,\, 24.6
   &  $\tilde u_1 W^-$\,,\,  33.2 & & \\
   & $\tilde u_1 W^-$\,,\,  38.9 & &   &   & & \\
\hline
$\tilde d_4 \to $  
   & $\tilde \chi^0_1 s$\,,\, 9.1 &  $\tilde \chi^0_1 b$\,,\, 6.3 &  $\tilde \chi^0_2 s$\,,\, 25.3
   & $\tilde \chi^0_1 d$\,,\, 2.3 & $\tilde \chi^0_2 d$\,,\, 31.7 &  $\tilde \chi^-_1 u$\,,\, 59.7
   \\ 
   & $\tilde \chi^-_1 u$\,,\, 2.1 &  $\tilde \chi^-_1 c$\,,\, 47.3 & $\tilde u_1 W^-$\,,\,  4.8 
   &  $\tilde \chi^-_1 c$\,,\, 3.0  & $\tilde \chi^-_2 u$\,,\, 2.3& \\
\hline
$\tilde d_5 \to $  
   & $\tilde \chi^0_1 d$\,,\, 2.3 & $\tilde \chi^0_2 d$\,,\, 31.7 &  $\tilde \chi^-_1 u$\,,\, 59.9
   & $\tilde \chi^0_1 s$\,,\, 2.2 &  $\tilde \chi^0_2 s$\,,\, 30.7 &  $\tilde \chi^-_1 u$\,,\, 2.9
   \\ 
   &  $\tilde \chi^-_1 c$\,,\, 2.8  & $\tilde \chi^-_2 u$\,,\, 2.3& 
  & $\tilde \chi^-_1 c$\,,\, 58.5 &  $\tilde \chi^-_2 c$\,,\, 2.3 &  \\
\hline
$\tilde d_6 \to $  
   & $\tilde \chi^0_1 s$\,,\, 3.1 & $\tilde \chi^0_2 s$\,,\, 30.6 &  $\tilde \chi^-_1 u$\,,\, 2.7
   & $\tilde \chi^0_1 s$\,,\, 19.7 &  $\tilde \chi^0_1 b$\,,\, 18.8 &  $\tilde \chi^0_3 b$\,,\, 2.9
   \\ 
  & $\tilde \chi^-_1 c$\,,\, 58.1 &  $\tilde \chi^-_2 c$\,,\, 2.4 & 
  & $\tilde \chi^0_4 b$\,,\, 2.9  & $\tilde \chi^-_2 t$\,,\, 5.8  & $\tilde g s$\,,\, 2.2  \\
  & & &
  & $\tilde g b$\,,\, 39.8 &$\tilde u_1 W^-$\,,\,  5.5 & \\
\hline
$\tilde g \to$ & $\tilde u_1 t$\,,\, 19.2 & $\tilde u_2 c$\,,\, 8.2 & $\tilde u_3 u$\,,\, 8.3 
               & $\tilde u_1 t$\,,\, 13.5 & $\tilde u_2 c$\,,\, 5.8 & $\tilde u_3 u$\,,\, 5.8 \\
               & $\tilde u_4 u$\,,\, 4.2 & $\tilde u_5 c$\,,\, 4.2 &  
               & $\tilde u_4 c$\,,\, 2.6 & $\tilde u_5 u$\,,\, 2.6 &  \\
          & $\tilde d_1 s$\,,\, 1.4 & $\tilde d_1 b$\,,\, 20.6 &  
          & $\tilde d_1 s$\,,\, 21.1 & $\tilde d_1 b$\,,\, 22.7  &  \\
          & $\tilde d_2 s$\,,\, 6.3 & $\tilde d_2 b$\,,\, 9.0 &  $\tilde d_3 d$\,,\, 8.3
          & $\tilde d_2 b$\,,\, 14.0 & &  $\tilde d_3 d$\,,\, 5.9 \\
          & $\tilde d_4 s$\,,\, 2.3 & $\tilde d_4 b$\,,\, 1.3 &  $\tilde d_6 s$\,,\, 2.8
          & $\tilde d_4 d$\,,\, 2.3 & $\tilde d_5 d$\,,\, 3.3  &  \\
\hline
\end{tabular}
%\end{center}
\end{table}
\begin{table}[hb]
\caption{Squark masses in GeV for SPS1a' and  our two points under study. The flavour
diagonal entries are according to SPS1a'. Note that $m_{\tilde u_2} \simeq  m_{\tilde u_3}$ 
and $m_{\tilde u_4} \simeq  m_{\tilde u_5}$.}
\label{tab:masses2}
\vskip 2mm \hskip4mm
%\begin{center}
\begin{tabular}{|c||c|c|c|c|c|c||c|c|c|c|}
\hline
 & $m_{\tilde d_1}$ & $m_{\tilde d_2}$ & $m_{\tilde d_3}$ 
 & $m_{\tilde d_4}$ & $m_{\tilde d_5}$ & $m_{\tilde d_6}$ 
 & $m_{\tilde u_1}$ & $m_{\tilde u_2}$ 
 & $m_{\tilde u_4}$ & $m_{\tilde u_6}$ \\
SPS1a' & 
   506 & 546 &  547 & 547 &    570 &  570 &
   367 & 547 &    565 &  586 \\
I.\,\,$\delta_{LL,23}=0.01,\delta_{d,RR,23}=0.1$ &
   503 & 525 &  547 & 569 &    570 &  570 &
   366 & 547 &     565 &  586 \\
II.\,\,$\delta_{LL,23}=0.04,\delta_{d,RR,23}=0.45$ &
   422 & 509 &  547 & 570 &  572 &  641 &
   366 & 547 &   565 & 587 \\
 \hline
\end{tabular}
%\end{center}
\end{table}

In the following we discuss the effect of the flavour mixing
parameters on the decay properties of squarks and gluinos. For the
discussion of the basic features we will first take two points in the
allowed region of Figure \ref{fig:2dregions}a). Afterwards  we discuss
various parameter dependencies. Note  that we often show only part of
the kinematically accessible final states and, thus, the branching
ratios shown do not sum necessarily up to 1. For the flavour diagonal
entries we will use the point SPS1a' but again similar features are
found for the other benchmark points as well. The main difference is
due to the kinematics.

The two study points chosen are characterized by $\delta_{LL,23}=0.01$
and $\delta_{D,RR23}=0.1$ (point I) and $\delta_{LL,23}=0.04$ and
$\delta_{D,RR23}=0.45$ (point II) respectively. Study point II is
characterized by large cancellations of the SUSY contributions to
$B$-physics observables.  In Table \ref{tab:BR} we give a summary of
the various branching ratios and in Table \ref{tab:masses2} we display
the masses. For comparison also the masses without flavour mixing
parameters are given.

The first feature to note is  that the bounds on $\delta_{LL}$ are
already so strong that there are small effects only on masses or
branching ratios due to this flavour-mixing parameter.\footnote{ Note
that this strong constraint would be weakened if more than two deltas
are active as shown in  the study point given in
ref.~\cite{Hurth:2003th} that contains such an example.}  This
implies that in our examples the masses and branching ratios of the
$u$-type squarks are hardly altered compared to the SPS1a'
point. Things are different for the $d$-type squarks and consequently
also for the gluino as the $\delta_{d,RR,23}$ parameter can be
sizable.

The relative size of the branching ratios in Table \ref{tab:BR} can be 
understood by the nature of  the various squarks mass eigenstates.
In point I one finds $\tilde d_1 \simeq \tilde b_L$ with a small
admixture of $\tilde b_R$,  $\tilde d_2 \simeq \tilde b_R$ with small
admixtures of $\tilde s_R$ and $\tilde b_L$, $\tilde d_3 \simeq \tilde
d_R$, $\tilde d_4 \simeq \tilde s_R$ with admixtures of $\tilde s_L$
and $\tilde b_R$, $\tilde d_5 \simeq \tilde d_L$ and $\tilde d_6
\simeq \tilde s_L$ with a small admixture of $\tilde s_R$. Thus,
larger flavour effects are visible in the decays of $\tilde d_2$ and
$\tilde d_4$ where the flavour violating decay branchings ratios
$\tilde d_2 \to \tilde\chi^0_1 s$ and $\tilde d_4 \to \tilde\chi^0_1
b$ are of the order of 10\%. This structure is also the reason for the
relative importance of the flavour violating decays of the gluino.  As
a side remark we note that the flavour violating decays of the first
generation squarks and of the 2nd/3rd generation squarks into the
first generation quarks are due to CKM quark mixing.

In point II the situation is more complicated  due to the larger flavour
mixing parameters.  With respect to the nature of the $d$-type squarks
we find that $\tilde d_1$ and $\tilde d_6$ are strongly mixed states
consisting mainly of $\tilde s_R$ and $\tilde b_R$ with a small
admixture of $\tilde b_L$ whereas the other states are mainly
electroweak eigenstates: $\tilde d_2 \simeq \tilde b_L$, $\tilde d_3
\simeq \tilde d_R$, $\tilde d_4 \simeq \tilde d_L$ and $\tilde d_5
\simeq \tilde s_L$. In this scenario the flavour violating final
states can even reach about  40\% in case of $\tilde d_2 \to
\tilde\chi^0_1 s$ and about 20\% for $\tilde d_6 \to \tilde\chi^0_1
b$. The differences for the gluino decays between these two points is
not only due to the different mixing in the $d$-squark sector but also
due to the different kinematics as can be seen from Table
\ref{tab:masses2}.

\begin{figure}[t]
 \unitlength 1mm
\begin{picture}(170,50)
\put(10,2){\mbox{\epsfig{figure=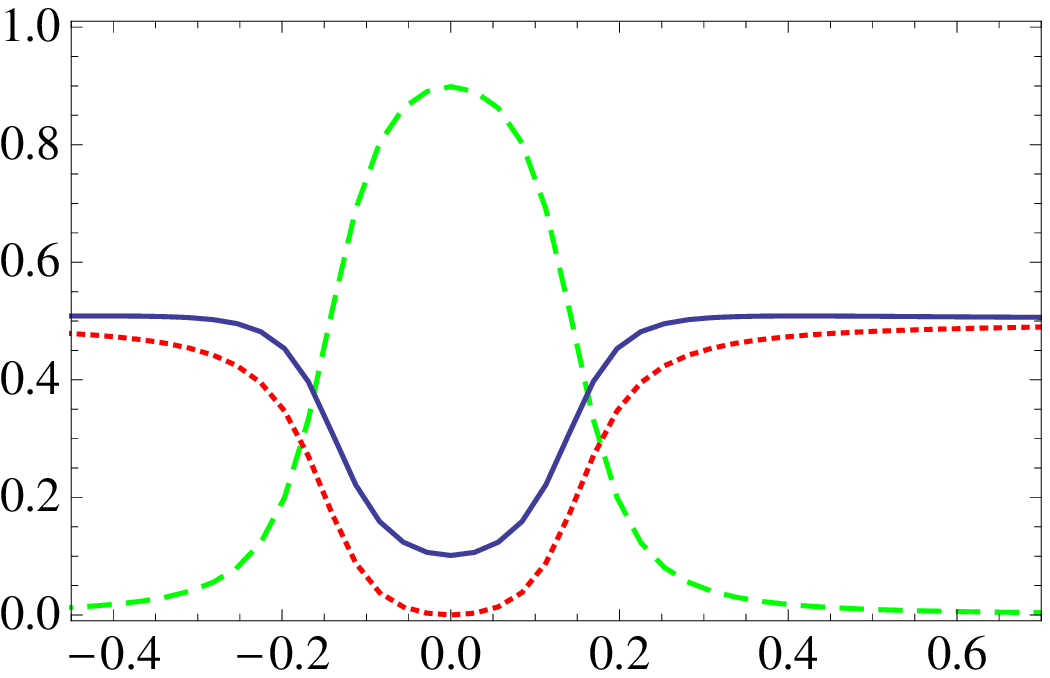,height=4.5cm,width=7cm}}}
\put(47,-2){\mbox{$\delta_{d,RR,23}$}}
\put(10,48){\mbox{a)}}
\put(90,2){\mbox{\epsfig{figure=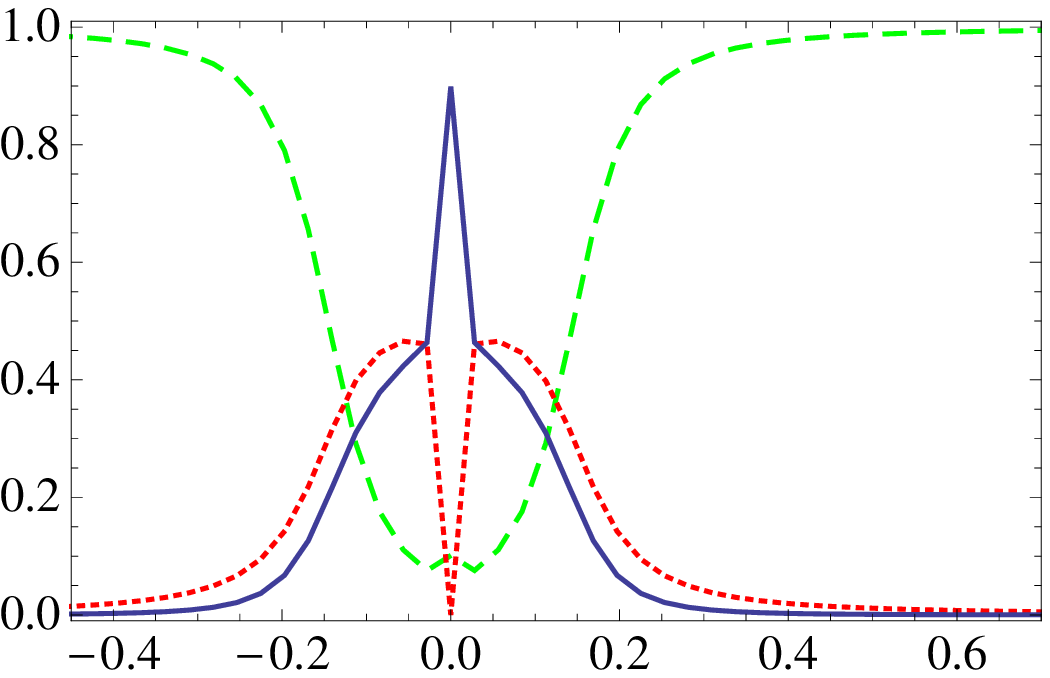,height=4.5cm,width=7cm}}}
\put(127,-2){\mbox{$\delta_{d,RR,23}$}}
\put(90,48){\mbox{b)}}
\end{picture}
\caption{ Composition of a) $\tilde d_{i=1}$ and b) $\tilde d_{i=2}$  as a function of  $\delta_{d,RR,23}$,
 the flavour diagonal
parameters are the ones of SPS1a'. The flavour basis is given by
$(\tilde d_L,  \tilde s_L,  \tilde b_L, \tilde d_R,  \tilde s_R,  \tilde b_R)$.
The lines  correspond to the final states:
dashed green line to $|R^d_{i,\tilde b_L}|^2$, 
dotted red line to $|R^d_{i,\tilde s_R}|^2$, and
full black line to $|R^d_{i,\tilde b_R}|^2$. }
\label{fig:RSdi}
\end{figure}

\begin{figure}[t]
 \unitlength 1mm
\begin{picture}(170,100)
\put(10,57){\mbox{\epsfig{figure=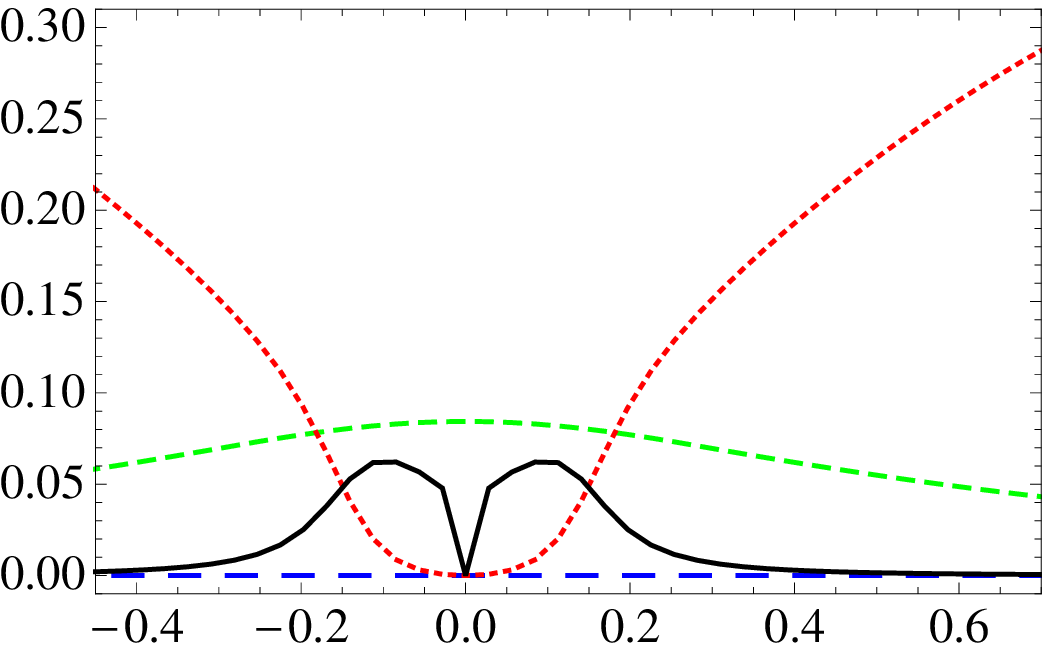,height=4.5cm,width=7cm}}}
\put(47,53){\mbox{$\delta_{d,RR,23}$}}
\put(10,103){\mbox{a)}}
\put(90,57){\mbox{\epsfig{figure=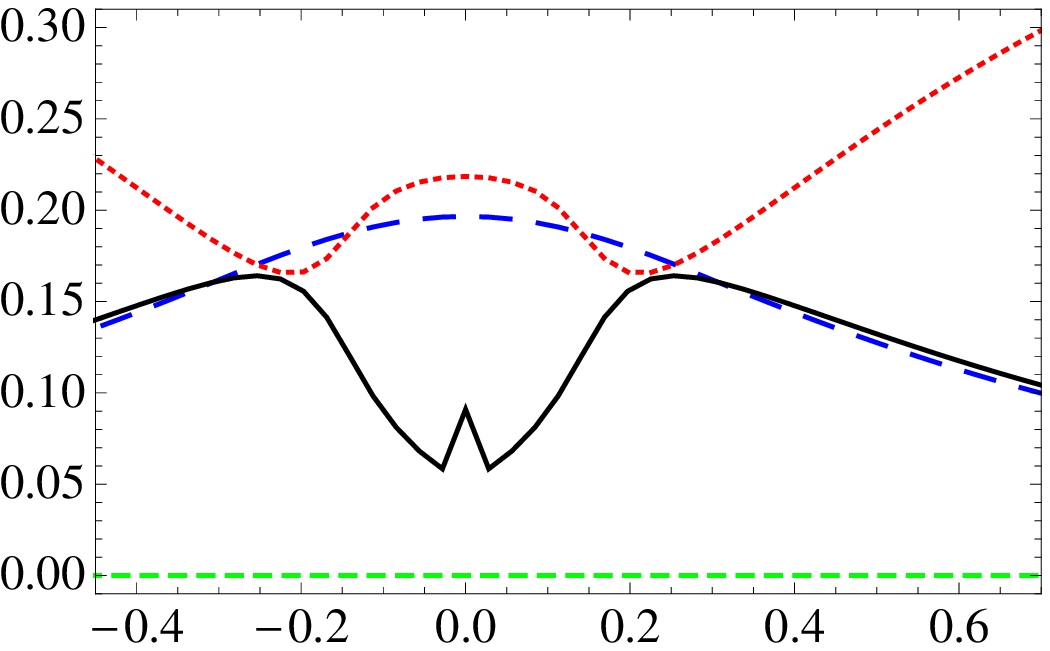,height=4.5cm,width=7cm}}}
\put(127,53){\mbox{$\delta_{d,RR,23}$}}
\put(90,103){\mbox{b)}}
\put(10,2){\mbox{\epsfig{figure=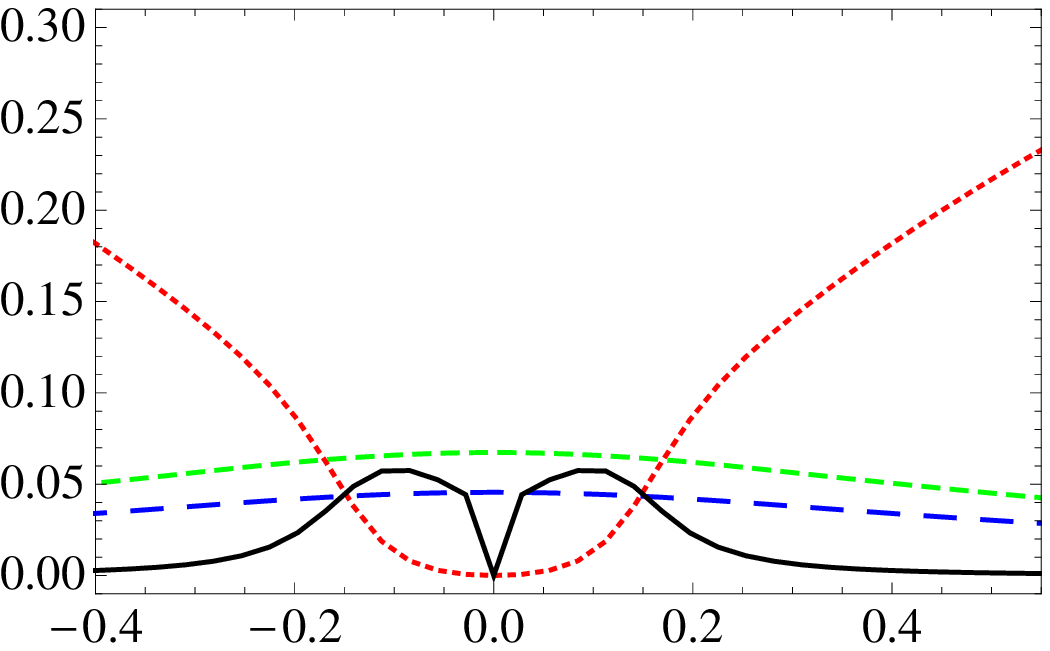,height=4.5cm,width=7cm}}}
\put(47,-2){\mbox{$\delta_{d,RR,23}$}}
\put(10,48){\mbox{c)}}
\put(90,2){\mbox{\epsfig{figure=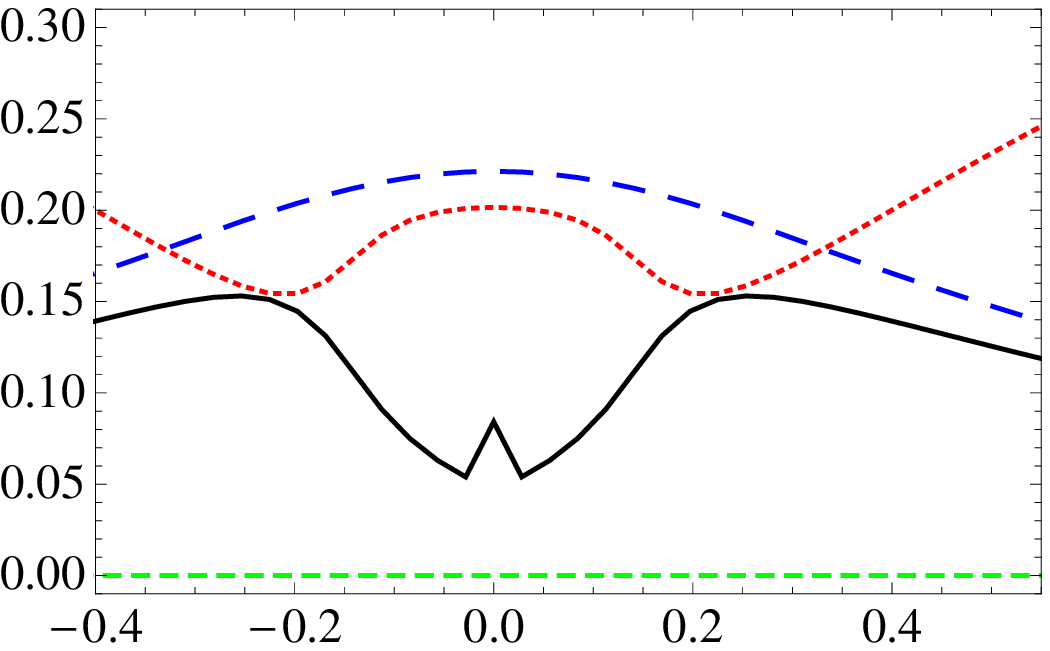,height=4.5cm,width=7cm}}}
\put(127,-2){\mbox{$\delta_{d,RR,23}$}}
\put(90,48){\mbox{d)}}
\end{picture}
\caption{$\tilde g$ decays as a function of $\delta_{d,RR,23}$, in a) and b)
all other $\delta$s are zero whereas in c) and d)  $\delta_{u,RR,23}=0.2$ in addition but all 
other  $\delta$ parameter  are zero.
The lines in a) and c) correspond to the final states $\tilde u_1 c$ (long dashed blue  line), 
$\tilde u_2 c$ (short dashed green line),  $\tilde d_1 s$ (dotted red line), and
$\tilde d_2 s$ (full black line); the lines in b) and d) correspond to the final states 
$\tilde u_1 t$ (long dashed blue line),    $\tilde u_2 t$ (short dashed green line), 
$\tilde d_1 b$ (dotted red line), and
$\tilde d_2 b$ (full black line).
 }
\label{fig:GluinoDecaysDRR}
\end{figure}
\begin{figure}[t]
 \unitlength 1mm
 \begin{picture}(170,50)
\put(10,2){\mbox{\epsfig{figure=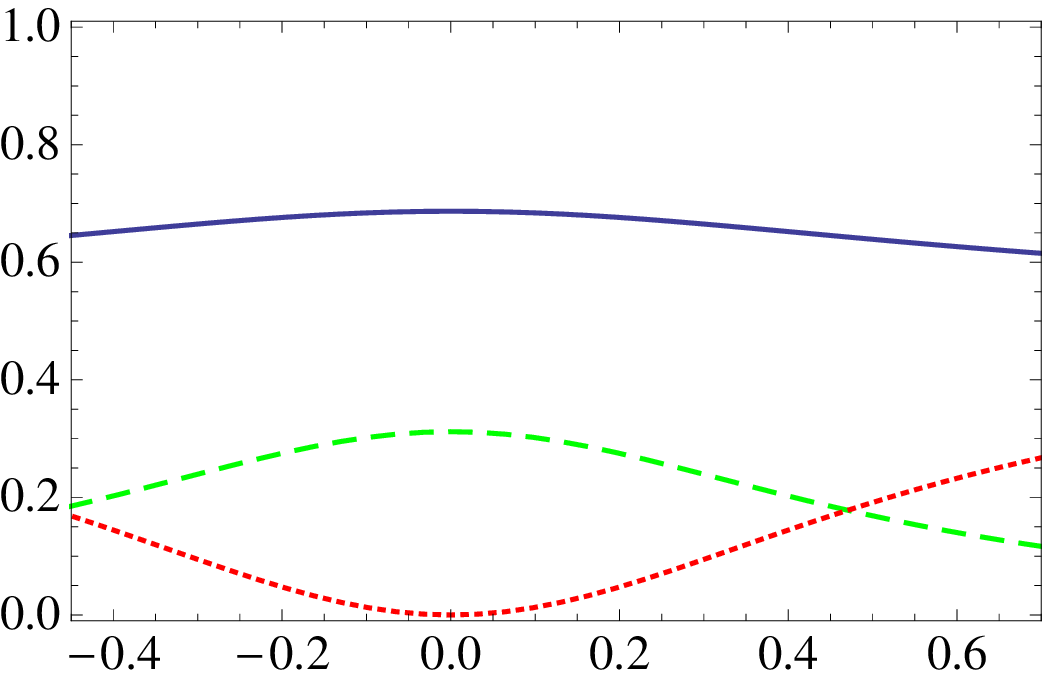,height=4.5cm,width=7cm}}}
\put(47,-2){\mbox{$\delta_{u,RR,23}$}}
\put(10,48){\mbox{a)}}
\put(90,2){\mbox{\epsfig{figure=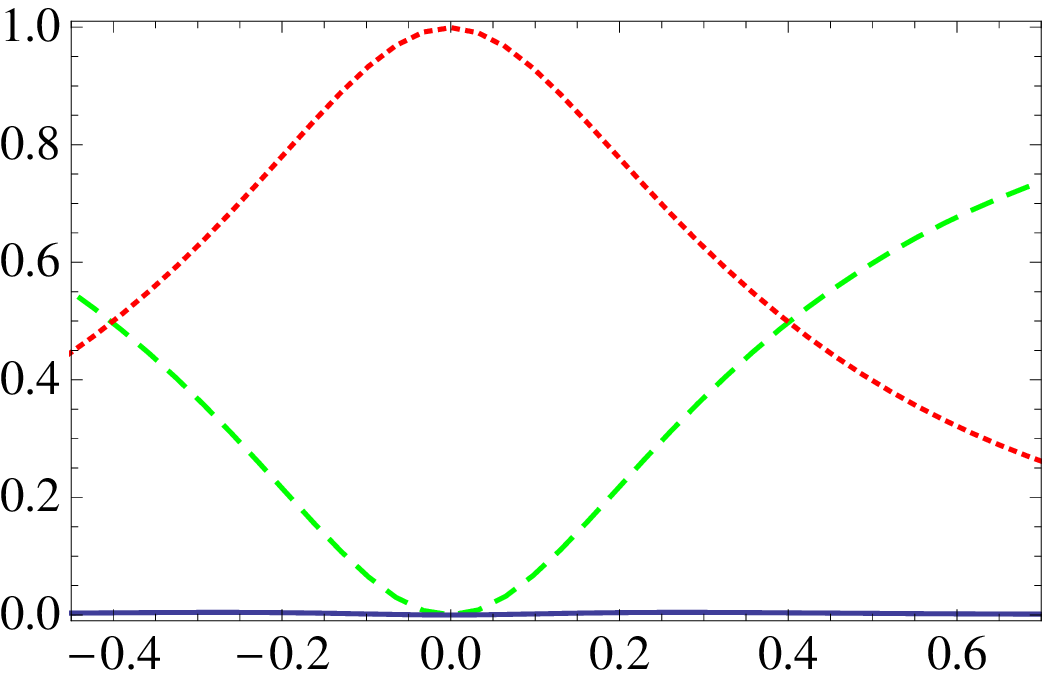,height=4.5cm,width=7cm}}}
\put(127,-2){\mbox{$\delta_{u,RR,23}$}}
\put(90,48){\mbox{b)}}
\end{picture}
\caption{ Composition of a) $\tilde u_{i=1}$ and b) $\tilde u_{i=2}$  as a function
 of  $\delta_{u,RR,23}$,  the flavour diagonal
parameters are the ones of SPS1a'. The flavour basis is given by
$(\tilde u_L,  \tilde c_L,  \tilde t_L, \tilde u_R,  \tilde c_R,  \tilde t_R)$.
The lines  correspond to the following final states:
dashed green line to $|R^u_{i,\tilde t_L}|^2$, 
dotted red line to $|R^u_{i,\tilde c_R}|^2$, and
full black line to $|R^u_{i,\tilde t_R}|^2$.}
\label{fig:RSui}
\end{figure}
\begin{figure}[th]
 \unitlength 1mm
\begin{picture}(170,50)
\put(10,2){\mbox{\epsfig{figure=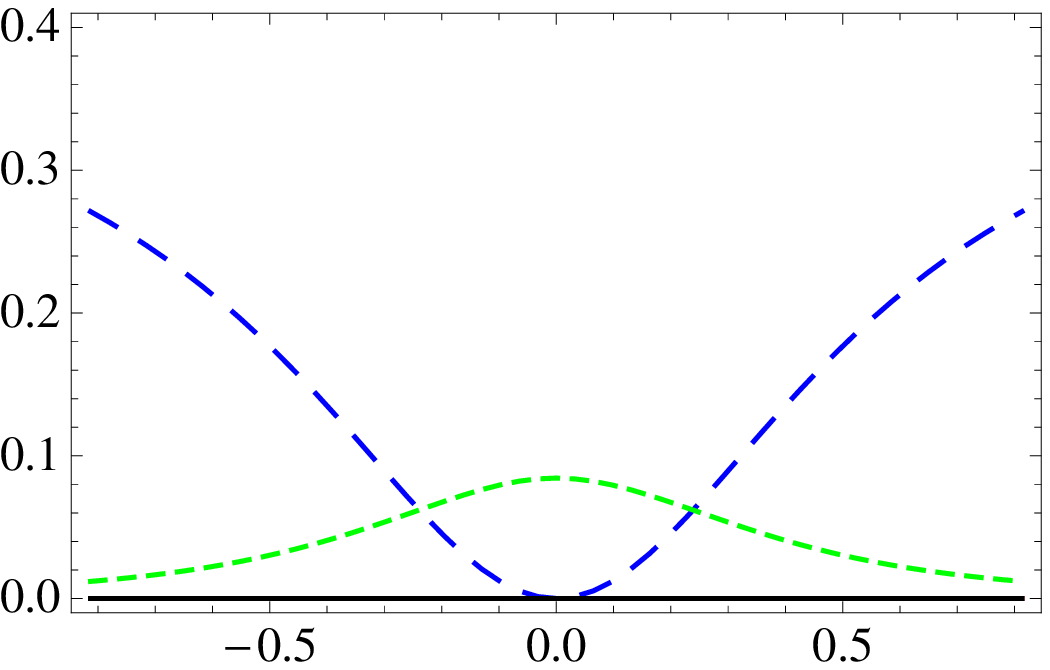,height=4.5cm,width=7cm}}}
\put(47,-2){\mbox{$\delta_{u,RR,23}$}}
\put(10,48){\mbox{a)}}
\put(90,2){\mbox{\epsfig{figure=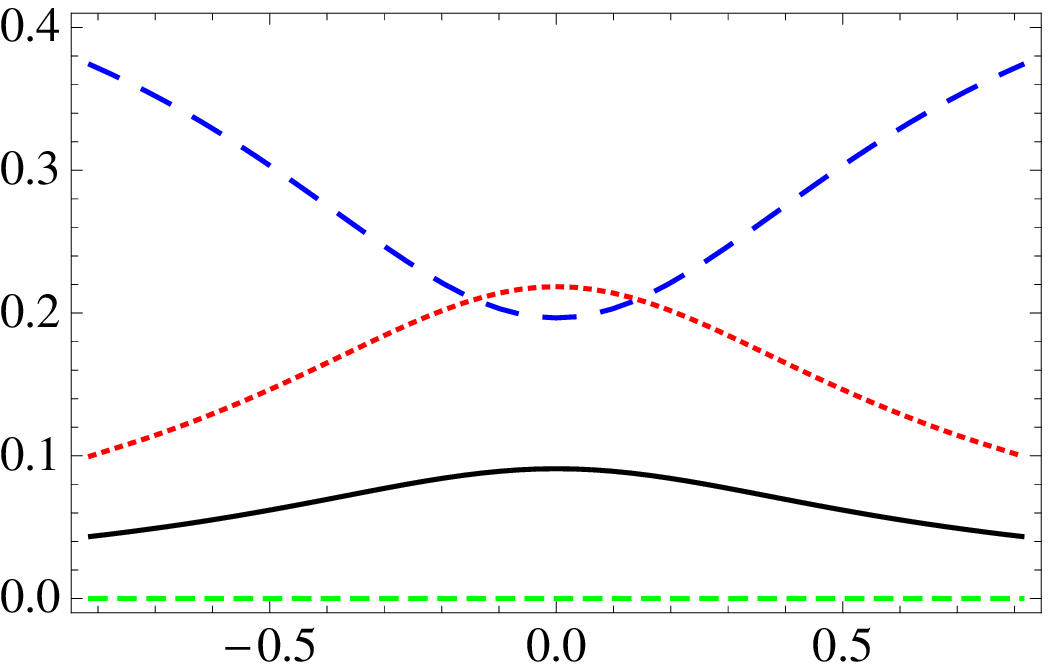,height=4.5cm,width=7cm}}}
\put(127,-2){\mbox{$\delta_{u,RR,23}$}}
\put(90,48){\mbox{b)}}
\end{picture}
\caption{$\tilde g$ decays as a function of $\delta_{u,RR,23}$ and all
 other $\delta$ parameters are zero.   The lines in a)  correspond to
the final states $\tilde u_1 c$ (long dashed blue line),  $\tilde u_2 c$  (short dashed green),  and 
$\tilde d_2 s$ (full black); the lines in b)  correspond to the
final states  $\tilde u_1 t$ (long dashed), $\tilde u_2 t$ (short dashed green), $\tilde d_1 b$ (dotted red), and $\tilde d_2 b$ (full black)}
\label{fig:GluinoDecaysURR}
\end{figure}
\begin{figure}[t]
 \unitlength 1mm
\begin{picture}(170,50)
\put(10,2){\mbox{\epsfig{figure=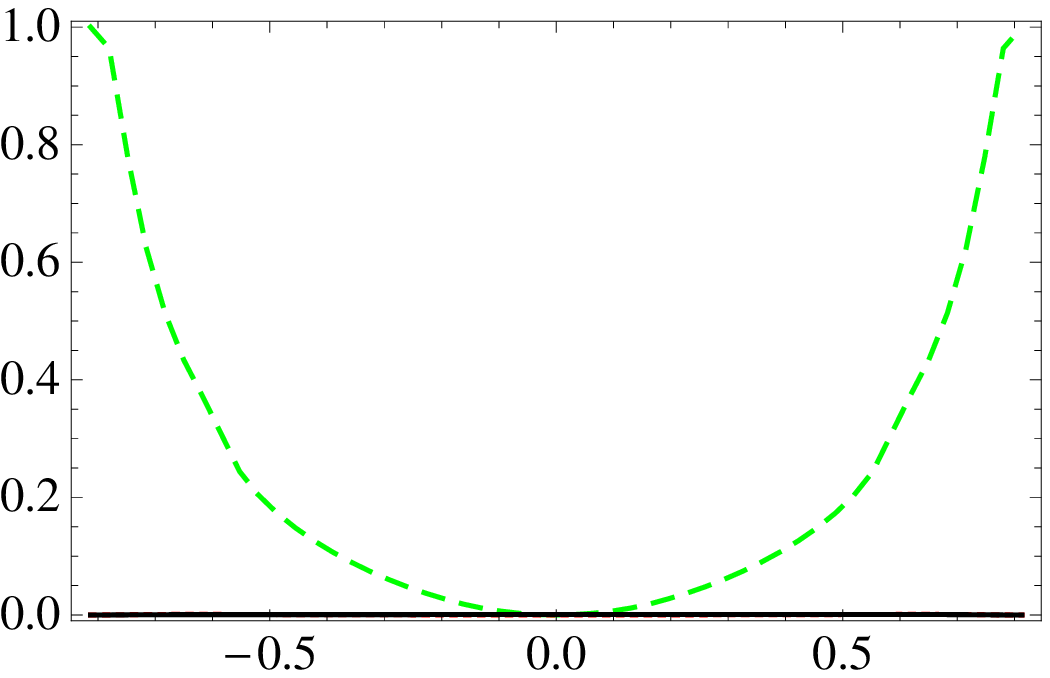,height=4.5cm,width=7cm}}}
\put(47,-2){\mbox{$\delta_{u,RR,23}$}}
\put(10,48){\mbox{a)}}
\put(90,2){\mbox{\epsfig{figure=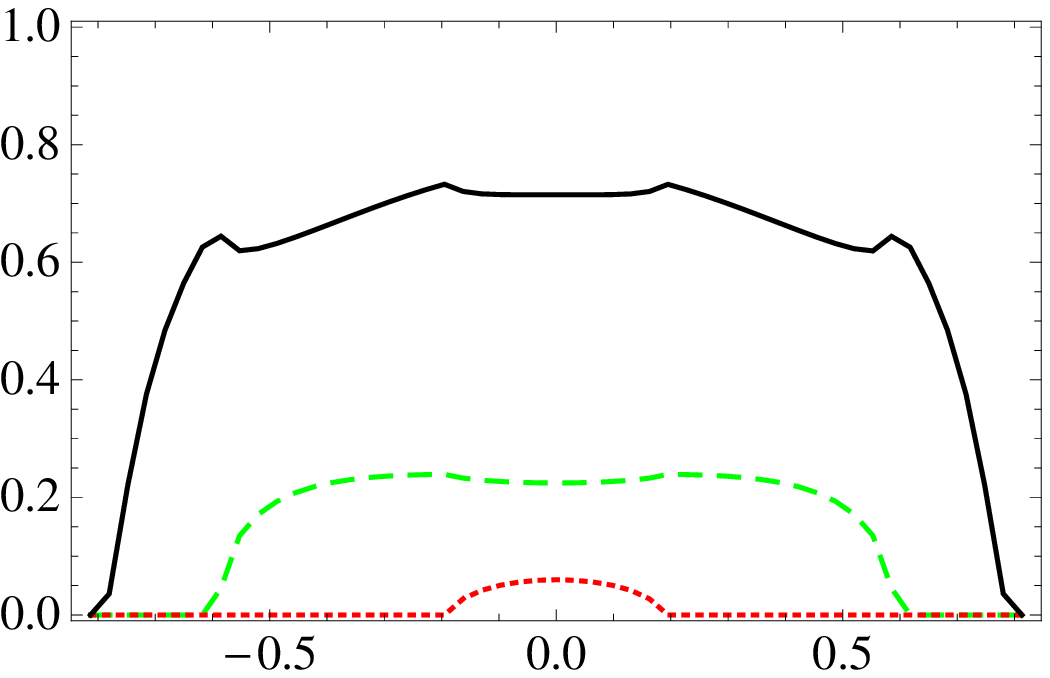,height=4.5cm,width=7cm}}}
\put(127,-2){\mbox{$\delta_{u,RR,23}$}}
\put(90,48){\mbox{b)}}
\end{picture}
\caption{$\tilde u_1$ decays as a function of  $\delta_{u,RR,23}$, the flavour diagonal
parameters are the ones of SPS1a'.  
The lines in a)  correspond to the final states $\tilde \chi_1^0 c$ (dashed green line), 
$\tilde \chi_2^0 c$ (dotted red), and 
$\tilde \chi^+_1 s$ (full black); the lines in b)  correspond to the final states
$\tilde \chi_1^0 t$ (dashed green), 
$\tilde \chi_2^0 t$ (dotted red), and
$\tilde \chi^+_1 b$. (full black).}
\label{fig:Sup1Decays}
\end{figure}\begin{figure}[t]
 \unitlength 1mm
\begin{picture}(170,50)
\put(10,2){\mbox{\epsfig{figure=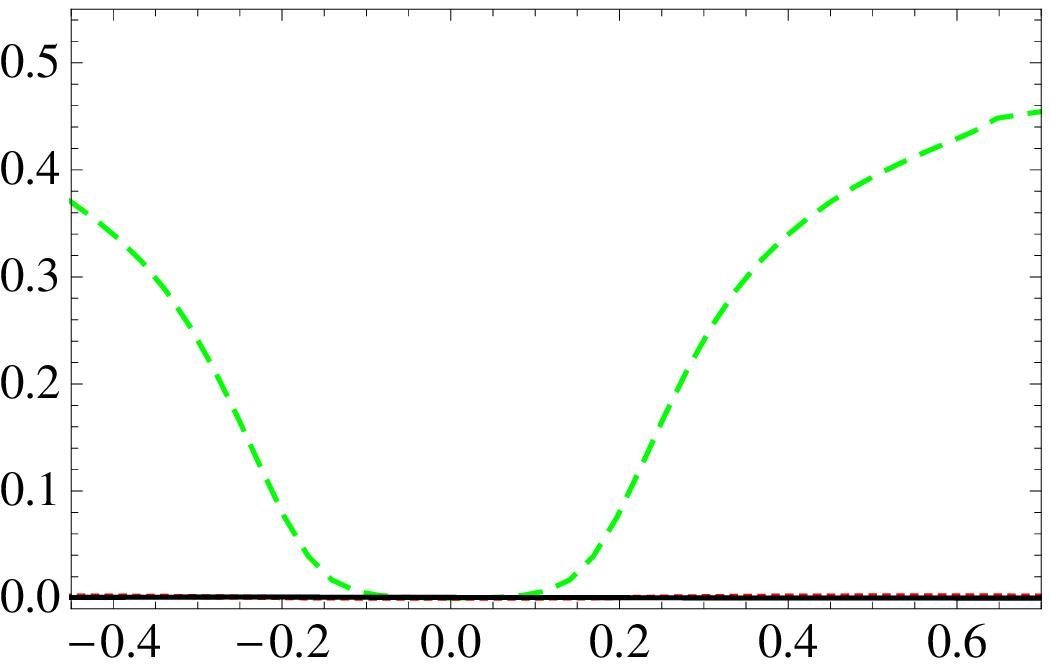,height=4.5cm,width=7cm}}}
\put(47,-2){\mbox{$\delta_{d,RR,23}$}}
\put(10,48){\mbox{a)}}
\put(90,2){\mbox{\epsfig{figure=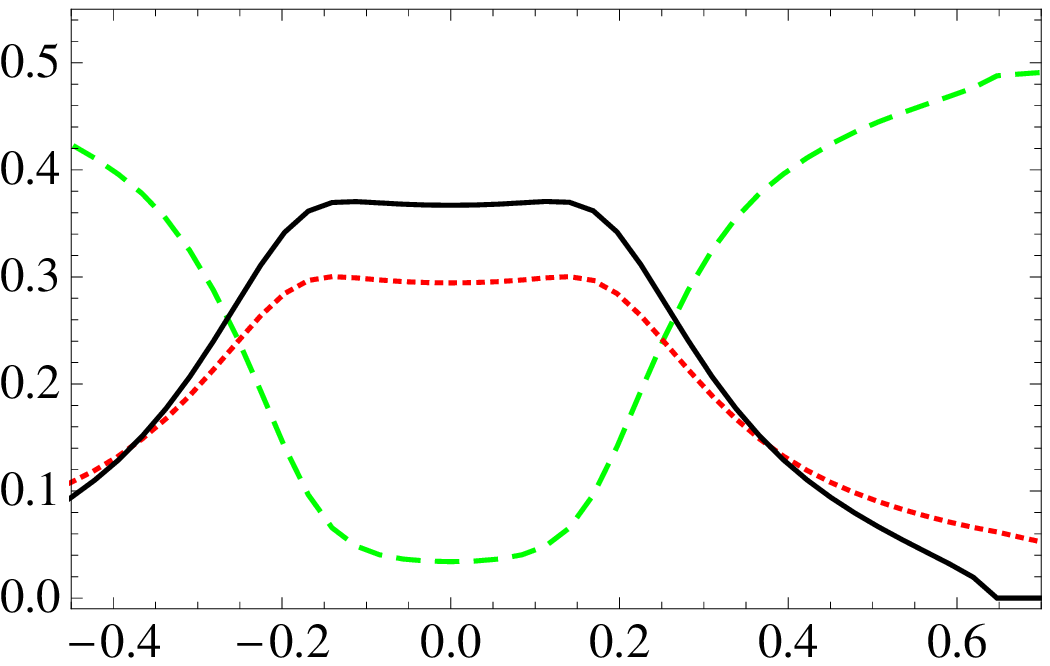,height=4.5cm,width=7cm}}}
\put(127,-2){\mbox{$\delta_{d,RR,23}$}}
\put(90,48){\mbox{b)}}
\end{picture}
\caption{$\tilde d_1$ decays as a function of  of  $\delta_{d,RR,23}$, the flavour diagonal
parameters are the ones of SPS1a'. 
The lines in a)  correspond to the final states $\tilde \chi_1^0 s$ (dashed green line), 
$\tilde \chi_2^0 s$ (dotted red), and $\tilde \chi^-_1 c$ (full black); the lines in b)  correspond to the final states 
$\tilde \chi_1^0 b$ (dashed green), 
$\tilde \chi_2^0 b$ (dotted red), and
$\tilde \chi^-_1 t$ (full black).}
\label{fig:Sdown1Decays}
\end{figure}

The effect of the flavour mixing parameters on gluino decays is strong
except for $\delta_{d,LR,ij}$. We first investigate how the nature
of the various squark states change when varying the flavour mixing entries.
In Figure \ref{fig:RSdi} we display the composition of the
two lightest $d$-squarks as a function of $\delta_{d,RR,23}$. The lightest
state is mainly a $\tilde b_L$ state for $|\delta_{d,RR,23}| \lsim 0.15$.
At $|\delta_{d,RR,23}| \lsim 0.2$ it is a strongly mixed state consisting of
$\tilde b_L$, $\tilde b_R$ and $\tilde s_R$ whereas for larger values
of $|\delta_{d,RR,23}|$ it is mainly an admixture of  
$\tilde b_R$ and $\tilde s_R$. The second state is mainly a $\tilde b_R$
state for $|\delta_{d,RR,23}| \lsim 0.05$.  The kinks around  0.05
stem   from the fact that the flavour mixing is about as large as the left-right
mixing in the sbottom sector. Around $|\delta_{d,RR,23}| \lsim 0.15$ the
$\tilde b_L$ component starts to dominate; for $|\delta_{d,RR,23}| \gsim 0.3$
this state is almost a pure $\tilde b_L$.

 This behaviour is 
reflected in the gluino decays as can be seen in \fig{GluinoDecaysDRR}
where we show the gluino decay branching ratios into the two lightest
$d$-squarks and into the two lightest $u$-squarks as a function of 
$\delta_{d,RR,23}$.   As already indicated in the
discussion above, the dependence is twofold. The larger the
off-diagonal elements are, the larger is the corresponding flavour
violating coupling; but also kinematical effects play an important
r\^ole, e.g.\ the larger $\delta_{d,RR,23}$  is,
the lighter $\tilde d_1$ gets  leading to an
increased phase space. The kinks and the crossing of the various lines
correspond exactly to the changes in the nature of the $d$-squarks.
Figure~\ref{fig:GluinoDecaysDRR}c shows in addition that there are scenarios
where one can have sizeable flavour violating decays into $d$-squarks and
into $u$-squarks at the same time.

Similar features can be observed for the case of large flavour mixing
in the $u$-squark sector. In Figure~\ref{fig:RSui} we display the
flavour content of the two lightest $u$-squarks as a function of 
$\delta_{u,RR,23}$. There are qualitative differences compared to the
$d$-squarks because (i) the left-right mixing in the top-squark
sector is significantly large and (ii) the lightest state $\tilde u_1$ in the
flavour conserving case is a mixture of $\tilde t_R $ and $\tilde t_L$ and 
$\tilde u_2  \simeq \tilde c_R$.  We see
that with increasing $\delta_{u,RR,23}$ not only the mixing between
$\tilde t_R$ and $\tilde c_R$ get larger as expected but that at the same
time the admixture of $\tilde t_L$ decreases quickly. The second state
is not, as one would naively expect, a strongly mixed state 
consisting of $\tilde t_R$ and $\tilde c_R$ but a mixture
between $ \tilde t_L$ and $\tilde c_R$ and only a negligible admixture
of $\tilde t_R$. The reason for this surprising feature is that the
left-right mixing in the top-squark sector is larger than
$\delta_{u,RR,23}$ for the complete range shown.

Again this is reflected in the gluino decays shown in \fig{GluinoDecaysURR}
where we display the gluino decay branching ratios as a function of
$\delta_{u,RR,23}$. Here the main flavour effect is the increase 
of the branching ratio into $\tilde u_1 c$ (long dashed blue line on 
the left side). The expected increase of $\tilde u_2 t$ does not take
place due to kinematics as $m_{\tilde u_2} + m_t > m_{\tilde g}$ for
most of the range shown. One would expect that the
branching ratio for $\tilde u_1 t$ should decrease for increasing
$|\delta_{u,RR,23}|$ as the top-squark components decrease. However, this decrease
is overcompensated by the increased kinematics as the mass of
$\tilde u_1$ drops to a value about 200 GeV.

The squarks stemming from the gluino decays will decay further and in
Figures \ref{fig:Sup1Decays} and \ref{fig:Sdown1Decays} we display the
corresponding branching ratios of the lightest $u$-squark and
$d$-squark, respectively. The behaviour of $\tilde u_1$ decays in
Figure \ref{fig:Sup1Decays} is a direct consequence of the above
described change of its nature with increasing
$|\delta_{u,RR,23}|$.
 But also its mass decrease again implies  that if $|\delta_{u,RR,23}|$
increases first the final state $\tilde \chi^0_2 t$ and then also
the final states
$\tilde \chi^+_1 b$ and $\tilde \chi^0_1 t$ get kinematically
forbidden.   The kinks in the various lines are a direct consequence of
these kinematical effects. Finally, for $|\delta_{u,RR,23}| \gsim 0.6$
only the final state $\tilde \chi^0_1 c$ is possible.  We have
checked that three and four body decay modes, such as $b W \tilde
\chi^0_1$ \cite{Porod:1996at}, $\tilde l \nu b$
\cite{Porod:1998yp,Djouadi:2000bx} or $b f \bar{f}' \tilde \chi^0_1$
\cite{Boehm:1999tr}, are suppressed and have branching ratios below 1\%.

In Figure \ref{fig:Sdown1Decays} we display the branching ratios of
$\tilde d_1$. The only sizeable flavour violating decay channel is
$\tilde \chi^0_1 s$ whereas $\tilde \chi^-_1 c$ is suppressed because
we consider here only flavour mixing between the $R$-squarks. The
above mentioned level crossing at $|\delta_{d,RR,23}| \simeq 0.15$ is
the main reason for the drop of the final states $\tilde \chi^0_2 b$
and $\tilde \chi^-_1 t$ besides kinematical effects.

\section{Impact for LHC and ILC}

\begin{figure}[th]
 \unitlength 1mm
\begin{picture}(17,50)
\put(10,2){\mbox{\epsfig{figure=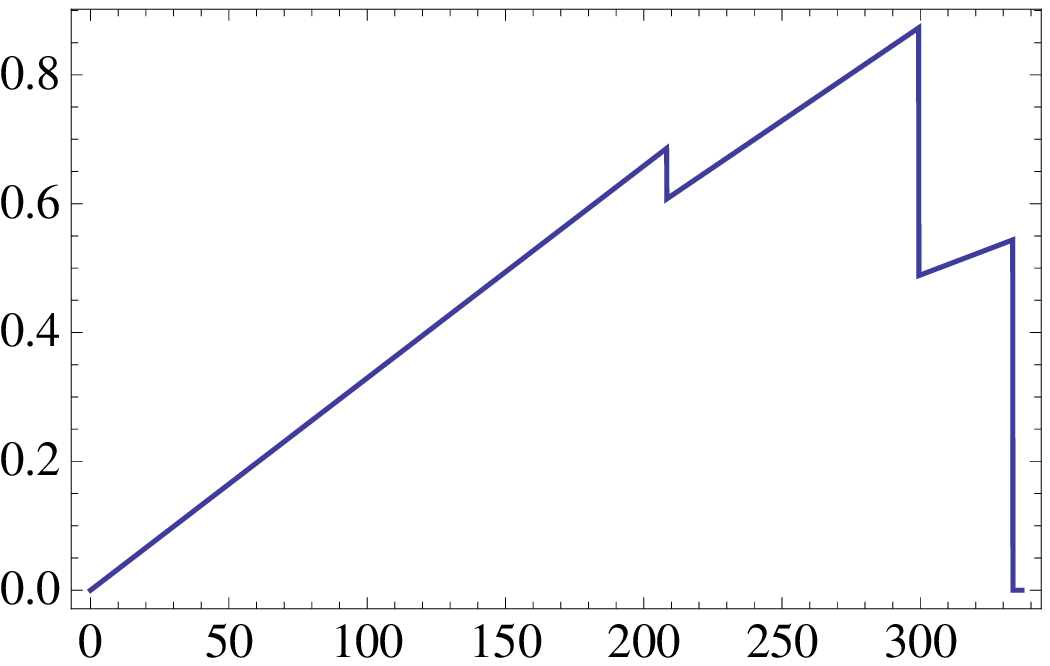,height=4.5cm,width=7cm}}}
\put(47,-2){\mbox{$m_{bb}$}}
\put(10,50){\mbox{{\bf a)} $10^4$ $d$(BR($\tilde g \to b \bar{b} \tilde \chi^0_1)/d m_{bb}$}}
\put(90,2){\mbox{\epsfig{figure=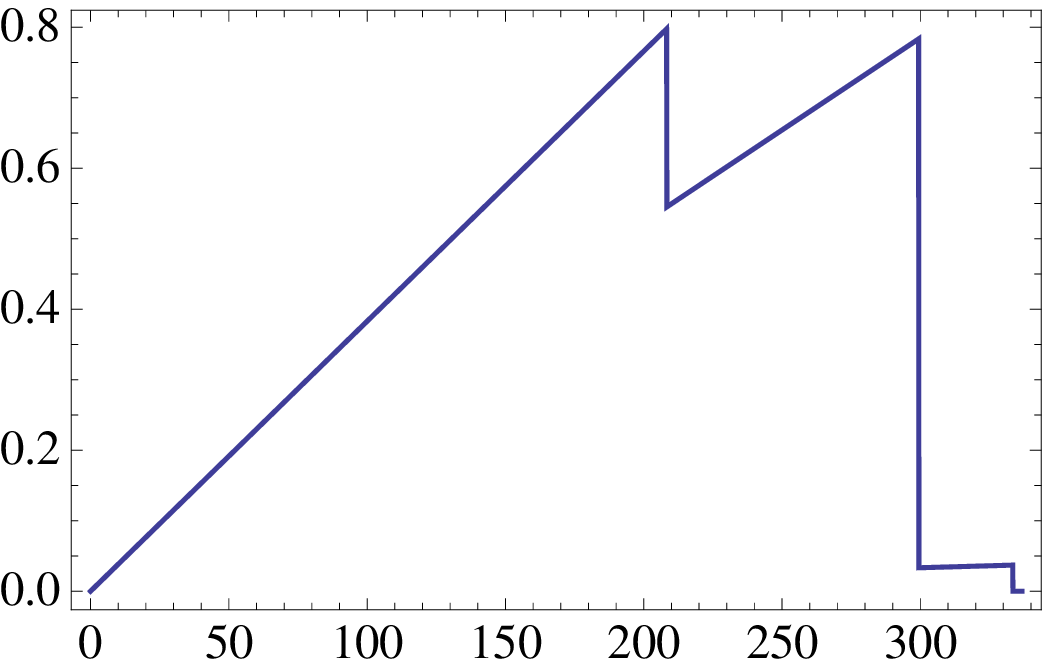,height=4.5cm,width=7cm}}}
\put(137,-2){\mbox{$m_{bs}$}}
\put(90,50){\mbox{{\bf b)} $10^4$ $d$(BR($\tilde g \to b s \tilde \chi^0_1)/d m_{bs}$}}
\end{picture}
\caption{Differential distributions $d$(BR($\tilde g \to b \bar{b} \tilde \chi^0_1)/d m_{bb}$
and $d$(BR($\tilde g \to b \bar{b} \tilde \chi^0_1)/d m_{bb}$
as a function of $m_{bb} = \sqrt{(p_b + p_{\bar{b}})^2}$ ($m_{bs}$) for point 1
defined in section \ref{sec:susydecays}. In b) the sum over the charges is shown:
BR($\tilde g \to b \bar{s} \tilde \chi^0_1)$ + BR$(\tilde g \to \bar{b} s \tilde \chi^0_1)$.
 }
\label{fig:bbbar}
\end{figure}

We have seen in Figure 1 that constraints due to low-energy 
observables lead to interesting structures in the parameter space and
the only certain parameter combinations are allowed in the case one
varies two parameters independently. Of course things are much more
complicated  once all off-diagonal elements are allowed to be non-zero
at the same time. In the end the crucial  question is  how  information
on flavour violating squark and gluino decays can be used to pin
down the underlying parameters. The answer to this question requires detailed
information on masses, production cross sections and branching ratios.

We have seen  that squarks and gluinos can have 
flavour changing decay modes of typically $O$(10)\% branching ratio. 
This clearly has an impact on the discovery
strategy of squarks and gluinos as well as on the measurement of the underlying
parameters at the LHC and a future international
linear collider (ILC). For example, in mSUGRA points without flavour mixing one
 finds usually that
the left-squarks of the first two generations as well as the right squarks
have similar masses. Large flavour mixing implies that there is a considerable
mass splitting as can be seen  in Table \ref{tab:masses2}. 
Therefore, the assumption
of almost degenerate masses should be reconsidered if sizable flavour changing
decays are discovered in squark and gluino decays.

An important part of the decay chains considered for SPS1a' and nearby points
are $\tilde g \to b \tilde b_j \to b \bar{b} \tilde \chi^0_k$ which are
used to determine the gluino mass as well as the sbottom masses or at least
their average value if these masses are close \cite{Branson:2001ak}.

In the latter analysis the existence of two $b$-jets has been assumed
stemming from this decay chain. In this case the two contributing
sbottoms would lead to two edges in the partial distribution
$d$(BR($\tilde g \to b \bar{b} \tilde \chi^0_1)/d m_{bb}$ where
$m_{bb}$ is the invariant mass of the two bottom quarks.  As can be
seen from Figure \ref{fig:bbbar} there are scenarios where more
squarks can contribute and consequently one finds a richer structure,
e.g.~three edges in the example shown corresponding to study point I.
Such a structure is either a clear sign of flavour violation or the
fact that the particle content of the MSSM needs to be extended.
Moreover, also the differential distribution of the final state $b s
\tilde \chi^0_1$ shows a similar structure where the edges occur at
the same places as in the $b\bar{b}$ spectrum but with different
relative heights. This gives a non-trivial cross-check on the
hypothesis of sizeable flavour mixing. Clearly a detailed Monte Carlo
study will be necessary to see with which precision one can extract
information on these edges. However, such a study is well beyond the
scope of this paper. Obvious difficulties will be combinatorics
because in general two gluinos or a gluino together with a squark will
be produced and, thus, there will be several jets stemming from light
quarks. However, one could take final states where one gluino decays
into $d$-type squarks and the second into stops or $c$-squarks. In the
second case effective charm tagging would be a crucial.

Similar conclusions hold for  the decay {chains 
$\tilde g\to c \tilde u_i \to c t \tilde \chi^0_1$ and 
$\tilde g\to t \tilde u_i \to c t \tilde \chi^0_1$}
 analyzed in \cite{KayNeu}  and also for the  
the variable $M^w_{tb}$ defined in 
\cite{Hisano:2003qu} which sums up final states  containing
$ t b \tilde \chi^+_1$. 

At an ILC the situation should be considerable easier: First one can
tune the center of mass energy so that one studies in principle the
states one after each other.  Secondly, one can polarize both,
electrons and positrons, \cite{Bartl:1997yi,MoortgatPick:2005cw} and
thus one influences the production rates for signal and
background. However, the ILC will be limited to energies in the one
TeV range and, thus, most likely only part of the squarks can be
explored at this machine. However, already this partial information
combined with LHC data can give useful information
\cite{Weiglein:2004hn}. Additional information can then be obtained at
a multi-TeV $e^+ e^-$ collider such as CLIC \cite{Accomando:2004sz}.

\section{Conclusions}

Flavour-violating low- and high-energy observables are governed by the
same parameters in supersymmetric models. A particular important
question is whether the soft SUSY breaking parameters can have
additional flavour structures beside the well-known CKM structure.
In this paper, we have analysed flavour-violating squark and gluino
decays in view of the present flavour data and have shown that they
can be typically of order of $10\%$ in the regions of
parameter space where no or only moderate cancellations between
different contributions to the low energy observables occur.
 If we allow for
larger  new physics contributions, e.g.~the same order as the SM
contributions, in the flavour observables, then even
flavour-violating branching ratios of up to $40\%$ are consistent with
the present data.  We have checked that this is a common feature for
a couple of SUSY benchmark points like SPS1a', $\gamma$, and $I''$.  We
have explicitly derived the pattern of flavour-violating decay modes from
the specific structure of the flavour-violating parameters including
kinematical constraints.  Finally, we have briefly analyzed the direct
consequences for the search of supersymmetric particles in specific
examples. The full exploitation of the necessary modifications in the
particle search calls for detailed Monte-Carlo analyses.

%\newpage 

\section*{Acknowledgments}
This work is supported by the European Network MRTN-CT-2006-035505 'HEPTOOLS'.
W.P.~is  is partially supported by the DFG, project Nr.\ PO 1337/1-1.

%\newpage 

\footnotesize
\begin{multicols}{2}

\end{multicols}
\end{document}